\documentclass[aps,reprint,amsmath,amssymb,superscriptaddress,preprintnumbers]{revtex4-2}

\usepackage{graphicx}
\usepackage{dcolumn}
\usepackage{bm}
\usepackage{braket}
\usepackage{hyperref}
\usepackage[export]{adjustbox}
\usepackage{mathtools}
\usepackage{dcolumn}

\usepackage{subfigure}

\usepackage{dsfont}

\usepackage{color}
\usepackage[dvipsnames]{xcolor}

\newcommand{\ch}[1]{\hat{c}_{#1}}

\newcommand{\ie}{\emph{i.e.}~}
\newcommand{\eg}{\emph{e.g.}~}

\newcommand{\beq}{\begin{equation}}
\newcommand{\eeq}{\end{equation}}
\newcommand{\bea}{\begin{eqnarray}}
\newcommand{\eea}{\end{eqnarray}}

\DeclareRobustCommand{\App}[1]{App.~\ref{#1}}
\DeclareRobustCommand{\Tab}[1]{Table~\ref{#1}}

\DeclareRobustCommand{\Fig}[1]{Fig.~\ref{#1}}

\DeclareRobustCommand{\Eq}[1]{Eq.~(\ref{#1})}

\usepackage{verbatim}

\begin{document}

\preprint{TUM-HEP-1531/24}

\title{A Universal Bound on QCD Axions from Supernovae}

\author{Konstantin Springmann}
\email{konstantin.springmann@weizmann.ac.il}
\affiliation{Technical University of Munich, TUM School of Natural Sciences, Physics Department, James-Franck-Str. 1, 85748 Garching, Germany}
\affiliation{Department of Particle Physics and Astrophysics,
Weizmann Institute of Science, Rehovot, Israel 7610001}

\author{Michael Stadlbauer}
\email{michael.stadlbauer@tum.de}
\affiliation{Technical University of Munich, TUM School of Natural Sciences, Physics Department, James-Franck-Str. 1, 85748 Garching, Germany}
\affiliation{Max Planck Institute for Physics, Boltzmannstr. 8, 85748 Garching, Germany}

\author{Stefan Stelzl}
\email{stefan.stelzl@epfl.ch}
\affiliation{Institute of Physics, Theoretical Particle Physics Laboratory, \'Ecole Polytechnique F\'ed\'erale de Lausanne, CH-1015 Lausanne, Switzerland}
\author{Andreas Weiler}
\email{andreas.weiler@tum.de}
\affiliation{Technical University of Munich, TUM School of Natural Sciences, Physics Department, James-Franck-Str. 1, 85748 Garching, Germany}

\date{\today}

\begin{abstract}
We identify a new production channel for QCD axions in supernova environments that contributes to axion emissivity for all models solving the strong CP problem. 
This channel arises at tree-level from a shift-symmetry-breaking operator constructed at next-to-leading order in Chiral Perturbation Theory. 
In scenarios where model-dependent derivative couplings to nucleons are absent, this sets the strongest model-independent constraint on the axion mass, improving on existing bounds by two orders of magnitude.

\end{abstract}

\maketitle


\section{Introduction}

In recent years, there has been increased interest in the physics of the QCD axion \cite{Peccei_1977a, Peccei_1977b, Wilczek_1978, Weinberg_1978}, supported by ongoing experimental efforts and new theoretical ideas.
While experiments increased their sensitivity enormously over the last decade, the strongest bounds still come from astrophysical probes, among which neutron star (NS) and supernova (SN) cooling are the most sensitive \cite{Raffelt:1987yt, Ellis:1987pk, Turner:1987by, Mayle:1987as, Mayle:1989yx, Brinkmann:1988vi,Raffelt:1991pw, Raffelt:1993ix, Janka:1995ir, Raffelt:1998pa, Sedrakian:2000kc, vanDalen:2003zw, Lykasov:2008yz,Chang:2018rso,Carenza:2019pxu,Guo:2019cvs, Entem:2017gor, Bartl:2016vwk,Buschmann:2021juv, Raffelt:1996wa,Caputo:2024oqc}.

In this work, we present a new production mechanism that is present for all axion models that solve the strong CP problem.
In particular, we investigate the shift-symmetry breaking operator that induces the nucleon electric dipole moment (EDM) operator at one-loop and point out that such an operator also leads to a production channel at tree-level.

We consider the pseudo-Nambu-Goldstone boson field $a(x)$ of a global axial $U(1)_\text{PQ}$, with coupling to quarks and gluons
\begin{equation} \label{eq:axion_GGdual}
    \mathcal{L}\supset\frac{g_s^2}{32\pi^2}\frac{a}{f_a}G_{\mu\nu}^a\tilde{G}_a^{\mu\nu}+\frac{\partial_{\mu} a}{2 f_{a}} J^{\mu}_{\mathrm{PQ,0}},
\end{equation}
where $G_{\mu\nu}^a$ is the field strength tensor of $SU(3)_c$, $f_a$ is the axion decay constant and $J^{\mu}_{\mathrm{PQ,0}}=\sum_{q} c_{q}^{0} \bar{q} \gamma^{\mu} \gamma_{5} q$ is the axial quark current, where $c_q^0 $ are model-dependent derivative couplings.
The $U(1)_\text{PQ}$ is anomalous, manifested in the fact that non-perturbative effects explicitly break the shift symmetry. 
At low energies, it is useful to apply an axial transformation on the light quark fields, \( q \to \exp\left(i a \gamma_5 Q_a / 2 f_a \right) q \), with $Q_a=M_q^{-1}/\operatorname{Tr}M_q^{-1}$ which removes tree-level axion-pion mixing. Note that here $M_q$ is the two flavor quark mass matrix.
This transformation eliminates the coupling to gluons described in Eq.~\eqref{eq:axion_GGdual} and produces a non-derivative coupling to the SM quarks,
\begin{equation}
    \label{eq:axion_quark_Lag}
    \mathcal{L} \supset- \left(\bar{q}_{L} M_{a} q_{R}+\text {h.c.}\right), \quad M_{a} \equiv e^{\frac{i a }{2 f_{a}}Q_{a}} M_q e^{\frac{i a }{2 f_{a}}Q_{a}}.
\end{equation}
Note that shift-symmetric couplings are also generated by the axial rotation.
These couplings are sensitive to the UV completion of the QCD axion and, as such, are model-dependent.
A particular choice of the model-dependent constants $c_{q}^0$ can lead to suppressed derivative couplings in the IR, for an explicit model realizing this, see \eg \cite{DiLuzio:2017ogq,Badziak:2023fsc}.
In contrast, the non-derivative couplings in \Eq{eq:axion_quark_Lag} are model-independent, as they are fixed by the requirement of the axion being a solution to the strong CP problem.

In this work, we are interested in axion production during SNe, where the dominant process is due to the couplings to nucleons. 
The dominant axion production process in typical SN environments is $NN\to NNa$, where $N=(p,n)^{\rm T}$ is the nucleon field. 
However, in the case of a large pion abundance, the process $\pi N \to a N$ can compete \cite{Carenza:2020cis}, and we will comment on it below.
It is possible to include (non-relativistic) baryons in the expansion of ChPT, known as Heavy Baryon ChPT (HBChPT), and hence to systematically derive the couplings of the QCD axion to nucleons and pions.
In a recent paper, we constructed the HBChPT Lagrangian for two light quark flavors up to NLO~\cite{Springmann:2024mjp}.
The couplings that get generated in the IR from Eq.~\eqref{eq:axion_quark_Lag} are found to be
\begin{equation} \label{eq:lag_non_pert}
    \mathcal{L}^{(2)}_{\pi N} \supset \hat{c}_1\bar{N} \braket{\chi_{+}} N+ \hat{c}_5 \bar{N} \tilde{\chi}_{+} N,
\end{equation}
where $\chi_{\pm}=u^{\dagger}\chi u^{\dagger} \pm u\chi^{\dagger}u,$ with $\chi =2 B {M}_{a}^\dagger$, $B$ is related to the quark condensate and $u = \exp(i \pi^{a} \tau^{a}/2 f_{\pi})$, with $\pi^a$ the pion fields and Pauli matrices $\tau^a$.
Here, we split $\chi_{\pm}$ into its flavor trace and traceless part, denoted by $\braket{\chi_\pm}$ and $\tilde{\chi}_{\pm}$, respectively. 
The low energy constants $\ch1$ and $\ch5$ can be found in \Tab{tab:constants}.

The first term in \Eq{eq:lag_non_pert} is negligible for particle production since it results in an axion-nucleon coupling involving at least two axions, which is thus suppressed by $1/f_a^2$.
\footnote{Note that this term has been extensively used in the literature as it can lead to the sourcing of light scalar fields in dense and large objects \cite{Hook:2017psm,Balkin:2020dsr,Balkin:2021zfd,Balkin:2021wea,Zhang:2021mks,Balkin:2022qer,Balkin:2023xtr,Gomez-Banon:2024oux}.}

This work focuses on the second term in Eq.~\eqref{eq:lag_non_pert}, which is isospin breaking and thus linear in the axion and the pion.
At one-loop this term gives rise to the EDM operator in \Eq{eq:ax_EDM_op} and at the tree level to a new axion production channel, see~\Fig{fig:c5_SN_diag}.
Compared to the derivative couplings of the axion, this production channel is suppressed: the operator arises at NLO in ChPT, and as it is isospin breaking, its coefficient is small.

However, as we will show, it is the dominant contribution to a model-independent bound on the axion \ie it provides the dominant contribution to the luminosity in models where derivative couplings are suppressed, such as the astrophobic axion \cite{DiLuzio:2017ogq}. 
This newly identified production channel establishes a bound that is two orders of magnitude more stringent than previously reported in the literature~\cite{Lucente:2022vuo}.

\section{Axion Production Channels} \label{sec:axion_channels}

In the absence of direct axion-nucleon coupling in $J^\mu_{\rm PQ}$, \eg the astrophobic axion, the leading order axion production via one-pion exchange \cite{Raffelt:1987yt, Ellis:1987pk, Turner:1987by, Mayle:1987as, Mayle:1989yx, Brinkmann:1988vi} is absent.
In \cite{Lucente:2022vuo}, it has been observed that QCD axions can be produced in supernovae via $NN \to NN a$ processes involving photon exchange, or through Compton-like scattering processes, $N\gamma \to N a$. 
These interactions are induced by the one-loop suppressed axion EDM operator, given by
\begin{equation} \label{eq:ax_EDM_op}
    \mathcal{L}_a^{\mathrm{EDM}}=-\frac{i}{2} \frac{C_{a N \gamma}}{m_N} \frac{a}{f_a} \bar{N} \gamma_5 \sigma_{\mu \nu} N F^{\mu \nu},
\end{equation}
where $\sigma^{\mu\nu} = \frac{i}{2} \left[ \gamma^\mu, \gamma^\nu \right]$, $m_N$ denotes the nucleon mass, $F^{\mu\nu}$ is the photon field strength tensor, and $|C_{a N \gamma}| \simeq 10^{-3}$ as determined in the seminal work of \cite{Crewther:1979pi}, see also \cite{Pospelov:1999ha}.

\begin{figure}[ht] 
  \centering
\begin{subfigure}
\centering
    \includegraphics[width=0.43\linewidth]{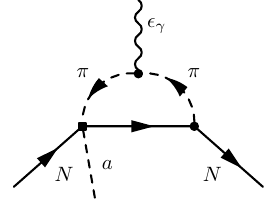}
\end{subfigure}
\begin{subfigure}
\centering
    \includegraphics[width=0.43\linewidth]{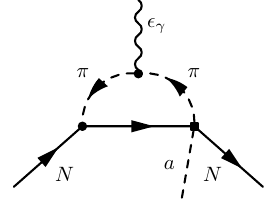}
\end{subfigure}
 \caption[]{EDM operator induced at 1-loop via the isospin breaking coupling $\ch5$.}
  \label{fig:oneLoop}
\end{figure}

This operator is induced at one-loop from the term $\propto \hat{c}_5$ in \Eq{eq:lag_non_pert} \footnote{A similar term is the dominant production channel for light pseudo-scalars that couple to the quark mass matrix. 
On the other hand, if, unlike the QCD axion, a pseudo-scalar field couples directly to a quark EDM, the nucleon EDM operator \cite{Lucente:2022vuo} is generated at tree level and thus dominant.}.
Expanding Eq.~\eqref{eq:lag_non_pert} to leading order in $1/f_a$,\cite{Springmann:2024mjp}, we find
\begin{equation}
    \label{eq:c5_operator}
    \mathcal{L}^{(2)}_{\pi N} \supset-\hat{c}_5m_\pi^2\frac{4z}{(1+z)^2}\bar{N} \left( \frac{\pi^a a}{ f_\pi f_a} \right) \tau^a N,
\end{equation}
where $z=m_u/m_d$ and $m_\pi$ and $f_\pi$ are the pion mass and decay constant, respectively.
Together with the leading order ChPT $\pi\pi\gamma$- and $NN\pi$-vertices, one can construct the one-loop diagrams in \Fig{fig:oneLoop}, which at low energies can be matched to Eq.~\eqref{eq:ax_EDM_op}, see~\cite{Crewther:1979pi}. 

The smallness of $C_{aN\gamma}$ can be understood as follows: Using the ChPT power counting (see e.g.~\cite{Weinberg:1978kz,Weinberg:1990rz,Jenkins:1990jv,Weinberg:1991um,Weinberg:1992yk}), we can estimate \begin{equation}
    \frac{C_{aN\gamma}}{m_N}\sim \frac{m_\pi^2}{(4\pi f_\pi)^2}\hat{c}_5.
\end{equation}

\begin{figure}[ht] 
  \centering
    \includegraphics[width=0.49\linewidth]{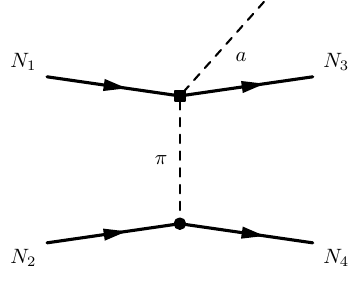}
 \caption[]{Tree-level LO axion production diagram via the isospin breaking Lagrangian Eq.~\eqref{eq:c5_operator}. The filled square is the vertex proportional to $\ch5$, while the filled circle is the LO HBChPT pion-nucleon vertex. This diagram is representative of 4 permutations where the axion vertex is connected to a nucleon line $N=(p,n)^{\rm T }$.}
  \label{fig:c5_SN_diag}
\end{figure}

In this work, we point out that \Eq{eq:c5_operator} directly contributes to a \emph{tree-level} contribution to the axion production in a SN, see \Fig{fig:c5_SN_diag}.
This bypasses the one-loop suppression, resulting in a constraint that is two orders of magnitude stronger.

Recently, we demonstrated in Ref.~\cite{Springmann:2024mjp} that higher order, as well as density corrections to the axion-nucleon interaction, can substantially alter the axion luminosity prediction from SNe, even reintroducing contributions that are otherwise suppressed in a vacuum.
While one might initially expect that large derivative couplings get re-introduced due to density corrections, this is not the case. 

The higher-order density corrections to the derivative axion-nucleon coupling are induced by two different types of diagrams. 
Some corrections are next-to-leading (NLO) order and proportional to the vacuum derivative axion-nucleon coupling, making them small, while others are suppressed due to their proportionality to the isospin-breaking coupling $c_5$ and their next-to-next-to-leading order nature.
This has been worked out recently in \cite{DiLuzio:2024vzg}. 
However, the production channel in \Fig{fig:c5_SN_diag} is independent of these density corrections and has been neglected so far.

\begin{figure}[h] 
  \centering
\begin{subfigure}
\centering
\includegraphics[scale=0.7,valign=c]{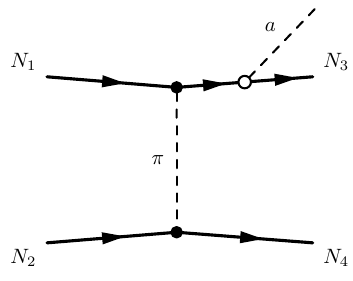} 
\end{subfigure}
\begin{subfigure}
\centering
\includegraphics[scale=0.7,valign=c]{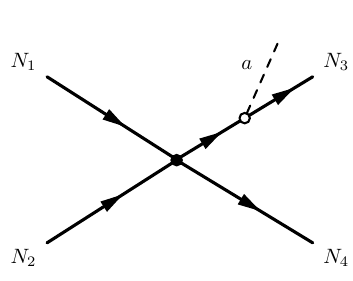} 
\end{subfigure}
\caption[]{Tree-level NLO axion production diagrams from the Lagrangian Eq.~\eqref{eq:nu2contribution}. The empty circle is the NLO axion-nucleon vertex. The LHS show the one-pion exchange while the RHS is the leading order contact diagram proportional to $C_S$, for details see \cite{Springmann:2024mjp}.
This diagram is representative of the 8 diagrams that arise from permutations of the axion vertex connected to a nucleon line $N=(p,n)^{\rm T }$.}
  \label{fig:nu2diagram}
\end{figure}

The next-to-leading order correction to \Fig{fig:c5_SN_diag} comes from diagrams shown in Fig.~\ref{fig:nu2diagram}, where the axion is attached to an external nucleon line by the model-independent operator
\begin{equation} \label{eq:nu2contribution}
    \mathcal{L}^{(3)}_{\pi N} \supset i \tilde{d}_{19} S^{\mu}\bar{N} \left[\partial_\mu, \braket{\chi_{-}} \right] N,
\end{equation}
where $S^{\mu}= \frac{i}{2} \gamma_5 \sigma^{\mu\nu} v_\nu$ is the nucleon spin operator and $v_\mu = (1,\vec{0})^T$~\footnote{Note that we are slightly deviating from the basis choice of \eg \cite{Fettes:2000gb} in that we split $\chi_-$ into its trace and its traceless part. }.
The axion appears in the trace contribution, and expanding to leading order, one finds 
\begin{equation}
    \braket{\chi_-}=im_\pi^2\frac{8z}{(1+z)^2}\frac{a}{f_a}+\dots .
\end{equation}
In analogy to \cite{Vonk:2020zfh}, we assume that the undetermined low-energy constant $\tilde{d}_{19}$ is drawn from a superposition of two normal distributions with $\tilde{d}_{19}=\pm 0.5(5)~ \text{GeV}^{-2}$.
This is estimated using naive-dimensional-analysis and matches the measured magnitude of the traceless version of the operator in \Eq{eq:nu2contribution}.
The corrections coming from these diagrams are suppressed by one power of $k/\Lambda$ compared to the leading order contribution. 
We include all corrections to this order, and estimate the effect of the truncation of the ChPT expansion for next-to-next-to leading order in analogy to \cite{Springmann:2024mjp}.
We neglect phenomenological effects taken into account in \cite{Chang:2018rso,Carenza:2019pxu,Carenza:2020cis} in order not to spoil the consistent systematics of our calculation.

In the following, we show the leading order squared matrix elements resulting from Fig.~\ref{fig:c5_SN_diag} for the two cases of interest: identical nucleons in the initial state ($nn$ or $pp$) and the mixed case ($np$).
For identical external nucleons, there are eight diagrams consisting of the two possible combinations of external states, \ie, $pp\to ppa$ and $nn\to nna$, each with $t-$ and $u-$channel, and two possibilities of attaching the axion vertex to the diagram.
They give rise to the squared matrix element
\begin{equation}
\begin{aligned}
& \hspace{-3mm} \left\lvert\mathcal{M}_{ nn/pp}\right\rvert^2 = \frac{4608 \, \hat{c}_5^2 \, g_A^2 \, m_N^4 \, m_\pi^4 }{f_a^2 \, f_\pi^4 }\frac{z^2}{(z + 1)^4}\,\,\times\\
& \hspace{-3mm} \Bigg[ \frac{\mathbf{k}^2}{\left( \mathbf{k}^2 + m_\pi^2 \right)^2}+\frac{\mathbf{l}^2}{\left( \mathbf{l}^2 + m_\pi^2 \right)^2}-\frac{2\left(\mathbf{k} \cdot \mathbf{l} \right)}{\left( \mathbf{k}^2+ m_\pi^2 \right) \left( \mathbf{l}^2 + m_\pi^2 \right)}\Bigg]. \\
\end{aligned}
\end{equation}
Here, $g_A$ is the axial isovector pion nucleon coupling, see \Tab{tab:constants}, and we defined $\mathbf{k} = \mathbf{p}_3 - \mathbf{p}_1$ and $\mathbf{l} = \mathbf{p}_4 - \mathbf{p}_1$, with $\mathbf{p}_i$ being the respective nucleon momenta.
For mixed nucleon initial and final states,  16 diagrams result from four external state permutations, contributions from both $t$- and $u$-channels, and two possible axion vertex attachments, leading to the squared matrix element,
\begin{equation}
   \left\lvert\mathcal{M}_{np}\right\rvert^2 = \frac{4608 \, \hat{c}_5^2 \, g_A^2 \, m_N^4 \, m_\pi^4 }{f_a^2 \, f_\pi^4}\frac{z^2}{\left(1+z\right)^4}\frac{\mathbf{k}^2}{\left( \mathbf{k}^2 + m_\pi^2 \right)^2}.
\end{equation}
Next, we discuss the NLO diagrams shown in \Fig{fig:nu2diagram}. 
The diagram in the left panel of \Fig{fig:nu2diagram} is of the same structure as the well known $NN\to NNa$ one-pion-exchange diagrams \cite{Brinkmann:1988vi,Carenza:2019pxu}. In contrast, the diagram in the right panel of \Fig{fig:nu2diagram} represents a nucleon contact interaction, with each featuring a different axion-nucleon vertex, see \Eq{eq:nu2contribution},
\begin{equation}
    8m_\pi^2\frac{z}{\left(1+z\right)^2} \tilde{d}_{19} \mathds{1} \frac{S \cdot p_a}{f_a},
\end{equation}
where $p_a$ is the axion momentum.
Note that this contribution is purely isoscalar. Although the NLO contribution is suppressed by a factor of $k/\Lambda$, drawing $\tilde{d}_{19}$ from its distribution results in it being larger than the LO contribution. This occurs because the coefficient $\hat{c}_5$ of the LO term is isospin-breaking and thus numerically small.

With the matrix elements determined, we proceed to calculate the axion production and emission rates relevant to SN explosions.

\section{Axion Luminosity and Bound}

The QCD axion provides an additional cooling channel for SN explosions.
The axion production rate or, equivalently, the emissivity $\dot{\varepsilon}_{a}$, \ie the emitted energy per volume and time, is given by 
\begin{equation}
    \begin{aligned}
\dot{\varepsilon}_{a}=&\int  \prod_{i=1}^4 d\Pi_{i} d \Pi_{a}(2 \pi)^{4}  \delta^{(4)}\left(\textstyle \sum_{i} p_i-p_{a}\right)  \\
& \quad \times S |\mathcal{M}|^{2} E_{a}f_{1} f_{2}\left(1-f_{3}\right)\left(1-f_{4}\right).
\end{aligned} \label{eq:emiss}
\end{equation}
Here, $S$ denotes a symmetry factor, $|\mathcal{M}|^2$ the matrix element squared, $d \Pi_{i}=d^{3} p_{i} /\left[(2 \pi)^{3} 2 E_{i}\right]$ the Lorentz invariant phase space measure, and \hbox{$f_i = (e^{[E_i(p_i) - \mu_i]/T} + 1)^{-1}$} is the Fermi-Dirac distribution of nucleon $i$. 
We perform this integration numerically using data from SN simulations, which include contributions from muons~\cite{Garching}, based on \cite{Bollig:2020xdr}.
\begin{figure}[b] 
  \centering
\includegraphics[width=1.\linewidth]{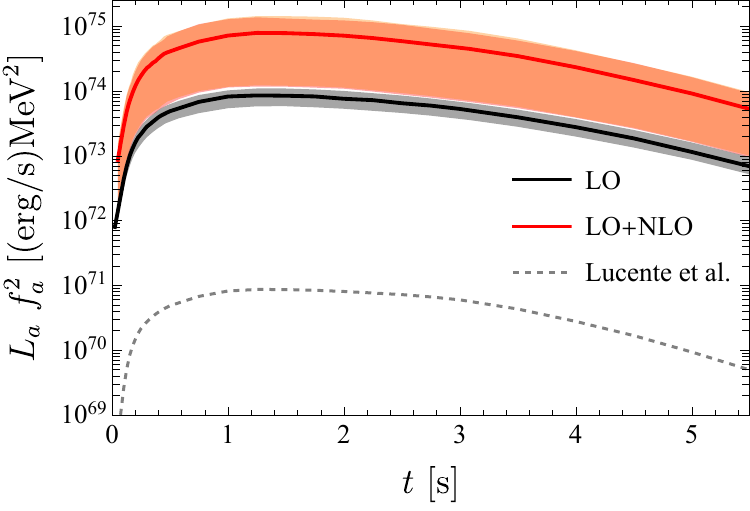}
 \caption[]{Axion luminosity $L_a f_a^2$ as a function of time for different contributions. 
In red, we show the LO and NLO contributions from this work, while the uncertainty arises from low-energy constants and ChPT expansion.
In black, we show the contribution from the LO tree-level diagram shown in \Fig{fig:c5_SN_diag}.
For comparison, we show the result of \cite{Lucente:2022vuo} for the same SN in gray dashed.}
\label{fig:Lumi}
\end{figure}
The luminosity, \ie the total energy emitted per unit time, is obtained by integrating the emissivity over the entire volume of the supernova
\begin{equation}
L_{a}=\int d r 4 \pi r^{2} \dot{\varepsilon}_{a}(r) \text{(lapse)}^2 (1 + 2v_r).
\end{equation}
We account for gravitational redshift and corrections from integrating the emissivity in the comoving frame, resulting in an additional radial-dependent factor of $\text{(lapse)}^2 (1 + 2v_r)$, as described in \cite{Rampp:2002bq,Marek:2005if,Garching}.

A bound on the QCD axion mass is obtained by imposing the condition that the axion luminosity is less than the neutrino luminosity $L_a\lesssim L_\nu$, during SN1987A \cite{Raffelt:2006cw}.
The input values for the Lagrangian parameters appearing in the calculation are summarized in \Tab{tab:constants}. They are obtained by matching to physical values at NLO as in \cite{Springmann:2024mjp}.
\begin{table}[t]
\begin{tabular}{|c|c|}
\hline
\multicolumn{1}{|l|}{Lagrangian parameter} & Value  \\ \hline
\hline
$g_A$ & $1.59(15)$ \\
$m_N$ & $852.9(19) ~\mathrm{MeV}$   \\
$f_\pi$ & $85.52(77) ~\mathrm{MeV}$  \\
$m_\pi$  & $141.7(75) ~\mathrm{MeV}$  \\
$z$    \cite{Workman:2022ynf}      & $0.474(65)$     \\
$\bar{\ell}_3$ \cite{FlavourLatticeAveragingGroupFLAG:2021npn} & 3.41(82) \\
$\bar{\ell}_4$ \cite{FlavourLatticeAveragingGroupFLAG:2021npn} & 4.40(28)  \\
$\ch1$  \cite{Hoferichter:2015tha}  & $-1.07(2) ~\mathrm{GeV}^{-1}$   \\
$\ch5$   \cite{Bernard:1996gq}  & $-0.09(1) ~\mathrm{GeV}^{-1}$  \\
$C_{S}$   \cite{Epelbaum:2001fm}  & $ -119(7) ~\mathrm{GeV}^{-2}$  \\
\hline
\end{tabular}
\caption{Numerical values of the bare Lagrangian parameters.} \label{tab:constants}
\end{table}
In \Fig{fig:Lumi} we show the QCD axion luminosity for LO and LO+NLO contributions and compare it to the result of \cite{Lucente:2022vuo}.
The axion luminosity, including LO+NLO contributions, exceeds the luminosity from \cite{Lucente:2022vuo} by a factor of around $10^3$ using the same SN data. A similar conclusion can be drawn from the LO contributions alone.
We apply the exclusion criterion that the axion luminosity at $t=1\,\text{s}$ must not exceed the neutrino luminosity at the same time. Based on our SN data, the neutrino luminosity is $L_\nu=5.5\times10^{52}\,\text{erg}\,\text{s}^{-1}$ \cite{Garching,Bollig:2020xdr}.

\begin{figure}[t] 
  \centering
\includegraphics[width=0.49\textwidth]{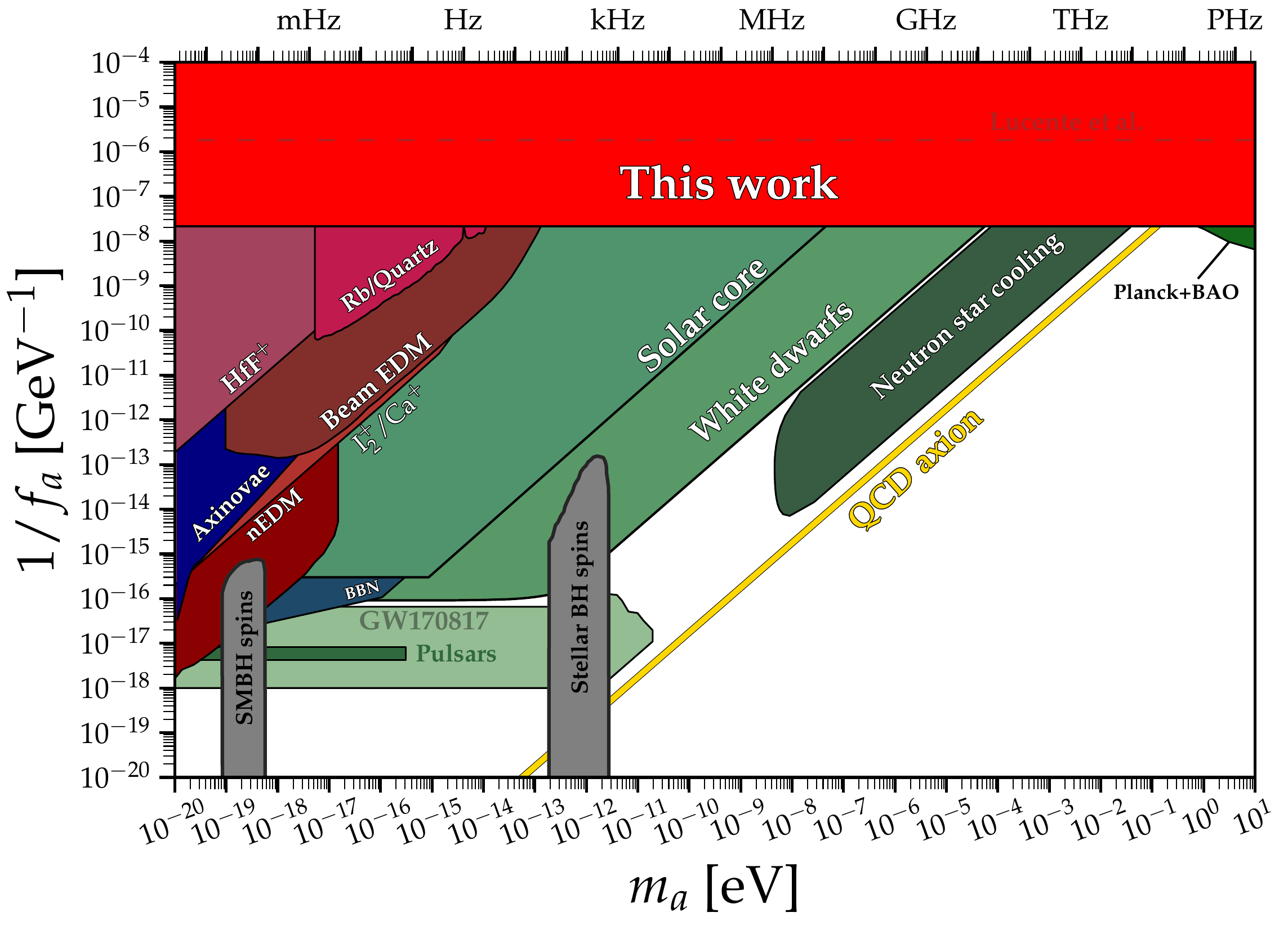}
 \caption[]{QCD axion parameter space adapted from \cite{AxionLimits}, $1/f_a$ as a function of $m_a$. The yellow line corresponds to the QCD axion. In red, we show the $68\%$ C.L. model-independent bound from SN 1987A obtained in this work. Other exclusions shown are from \cite{AxionLimits,Schulthess:2022pbp,Abel:2017rtm,Roussy:2020ily,Madge:2024aot,Fuchs:2024edo,JEDI:2022hxa,Zhang:2022ewz,Fox:2023xgx,Blum:2014vsa,Mehta:2020kwu,Baryakhtar:2020gao,Unal:2020jiy,Hook:2017psm,DiLuzio:2021pxd,Zhang:2021mks,Caloni:2022uya,Lucente:2022vuo,Balkin:2022qer,Badziak:2024szg,Gomez-Banon:2024oux,Hoof:2024quk}.}
  \label{fig:exclusion}
\end{figure}

This leads to the strongest model-independent bound on the axion decay constant
\begin{equation}\label{eq:FullBoundLumi}
    f_a >  1.1^{+0.4}_{-0.6} \times 10^8 \, \mathrm{GeV},\quad\text(68\%\,\,\text{C.L.})
\end{equation}
or equivalently on the QCD axion mass
\begin{equation}
    m_a < 5.3^{+7.3}_{-1.3} \times 10^{-2} \, \mathrm{eV}. \quad\text(68\%\,\,\text{C.L.})
\end{equation}
The error estimate originates from the uncertainty of the input parameters and the truncation of the chiral expansion at NLO.
We show the corresponding QCD axion parameter space, including our novel model-independent bound, and compare it to existing limits in \Fig{fig:exclusion}.

In \App{app:alternative_process} we investigate the alternative axion production process ($\pi N \to a N$) present assuming large pion chemical potentials \cite{Carenza:2020cis}. It is straightforward to estimate that similar considerations lead to a bound of $f_a \gtrsim 4 \times 10^7~\mathrm{GeV}$.

\section{Conclusions}

Supernova cooling provides one of the strongest bounds on the QCD axion.
In this work, we have shown that the non-derivative operator, responsible for inducing the nEDM operator at one-loop, also generates a tree-level axion production channel in supernova environments.
The coefficient of this operator is fully determined for any QCD axion that solves the strong CP problem.
In models where the QCD axion has small derivative couplings to nucleons, this production channel becomes the dominant mechanism.
Our results show that the axion luminosity from this channel exceeds previous estimates, which relied on the nEDM portal, by roughly three orders of magnitude. 
Comparison with the neutrino luminosity from SN 1987A yields the strongest model-independent constraint on the QCD axion mass, without requiring the axion to make up a significant fraction of dark matter.

In particular, we find at $68\%$ C.L.
\begin{equation}
    f_a >  1.1^{+0.4}_{-0.6} \times 10^8 \, \mathrm{GeV},\quad m_a < 5.3^{+7.3}_{-1.3} \times 10^{-2} \, \mathrm{eV}.
\end{equation}
These results demonstrate the effectiveness of a systematic ChPT approach in establishing model-independent constraints on the QCD axion, which even astrophobic axions cannot evade.


\begin{acknowledgements}
We would like to thank Thomas Janka and his group for providing us with supernova simulation data.
This work has been supported by the Collaborative Research Center SFB1258, the Munich Institute for \hbox{Astro-, Particle} and BioPhysics (MIAPbP), and by the Excellence Cluster ORIGINS, which is funded by the Deutsche Forschungsgemeinschaft (DFG, German Research Foundation) under Germany's Excellence Strategy – EXC 2094 – 390783311. The research of MS is partially supported by the International Max Planck Research School (IMPRS) on “Elementary Particle Physics”.
KS is supported by a research grant from Mr. and Mrs. George Zbeda.
SS is partially supported by the Swiss National Science Foundation under contract 200020-213104. SS thanks the CERN theory group for its hospitality.
\end{acknowledgements}

\appendix

\section{Alternative process} \label{app:alternative_process}

We comment on the role of alternative axion production processes.
If there is a non-negligible pion abundance in the SN, pion-axion conversion $N\pi\to N a$ can compete with the Bremsstrahlung process \cite{Carenza:2020cis}.
As this process depends on the details of the supernova, we do not combine the production channels, but for completeness, we estimate its importance in the case where the dominant axion coupling is $\propto \ch5$.

\begin{figure}[h] 
  \centering
\begin{subfigure}
  \centering
\includegraphics[width=0.48\linewidth]{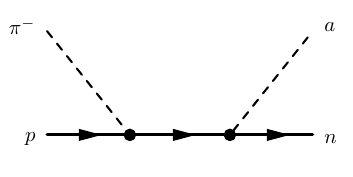}
\end{subfigure}
\begin{subfigure}
  \centering
\includegraphics[width=0.48\linewidth]{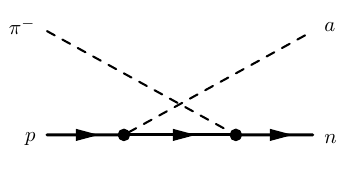}
\end{subfigure}
 \caption[]{Diagrams relevant for the axion production via pion conversion in SNe with model-dependent axion-nucleon couplings, exploited in \cite{Lucente:2022vuo}.}
  \label{fig:pion_conv}
\end{figure}
\begin{figure}[h] 
  \centering
\includegraphics[width=0.48\linewidth]{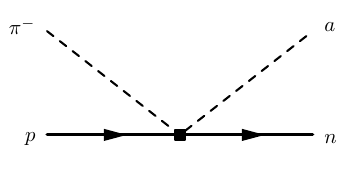}
 \caption[]{Diagram relevant for axion production in SNe via pion conversion from the operator \Eq{eq:c5_operator}.}
  \label{fig:pion_conv_c5}
\end{figure}

We first review the pion conversion process in the case of large derivative couplings \cite{Carenza:2020cis}, with the relevant diagrams shown in \Fig{fig:pion_conv}.
The squared matrix element is given by
\begin{equation} \label{eq:matrix_Carenza}
|\mathcal{M}|^2 = \frac{2 m_N^2 \mathbf{p}_\pi^{2}}{3 f_\pi^2 f_a^2} (c_{u-d}^2 g_A^4 + 2 c_{u+d}^2 g_0^2 g_A^2).
\end{equation}
If, however, derivative couplings are absent, the dominant process is again $\propto\ch5$ as shown in \Fig{fig:pion_conv_c5}. 
The squared matrix element in this case is
\begin{equation} \label{eq:matrix_c5}
|{\mathcal{M}}_{\ch5}|^2 = \frac{128 \ch5^2 m_N^2 m_\pi^4}{f_a^2 f_\pi^2} \frac{z^2}{(1+z)^4}.
\end{equation}
The phase space integrals, and thus the luminosity, are dominated for momenta $\mathbf{p}_\pi^2\sim m_\pi$, leading to
\begin{equation}
    |{\mathcal{M}}_{\ch5}|^2/|{\mathcal{M}}|^2 \simeq 42 \, \frac{\ch5^2 m_\pi^4}{\mathbf{p}_\pi^2} \xrightarrow[\mathbf{p}_\pi^2\to m_\pi^2]{} 6 \times 10^{-3}.
\end{equation} 
Rescaling the bound on $f_a$ of \cite{Carenza:2020cis} yields
\begin{equation} \label{eq:bound_4pt}
    (f_a)_{\ch5} \gtrsim 4 \times 10^7~\mathrm{GeV},
\end{equation}
of similar size as the Bremsstrahlung process. However, it heavily relies on the existence of a large pion abundance.

\section{Dependence on the Supernova Model}

While the model-independent supernova bound put forward in this paper is independent of the specific QCD axion model, it does depend on the specifics of the supernova model. 
In this appendix, we study some aspects of the supernova model dependence on the bound derived.

\begin{figure}[ht] 
  \centering
\includegraphics[width=1.\linewidth]{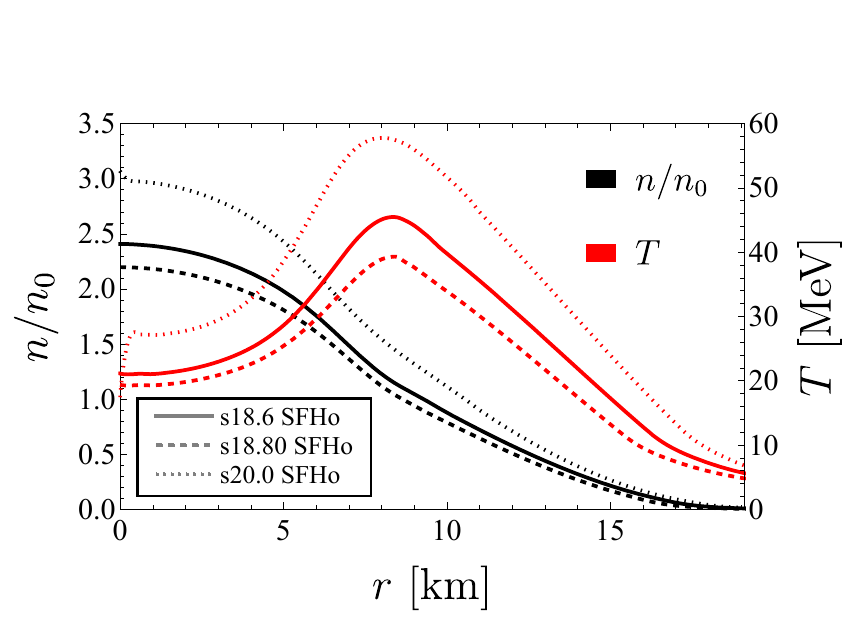}
\caption[]{Density and temperature profile taken from a supernova simulation of \cite{Garching} at $t_{pb}=1\,\mathrm{s}$ for the three supernova models discussed in this section.}
\label{fig:profile_model_dep}
\end{figure}

In the main text, we have studied the axion luminosity resulting from one specific supernova model based on the SFHo EOS that included muons with a progenitor star of a mass of $18.6 \,M_\odot$ from~\cite{Garching}. 
In this appendix, we extend the analysis by two different supernova models, based on the same EOS but different progenitor masses and simulations, and perform the same calculation, again at LO.
The SN simulations we take additionally into account have a $18.8\,M_\odot$ and $20\,M_\odot$ progenitor masses, both with the SFHo EOS and including muons, see~\cite{Garching}.
While the quantitative numerical results differ for each of the models by an order of one factor, qualitatively, the results are in agreement.

\begin{figure}[t]
  \centering
\includegraphics[width=1.\linewidth]{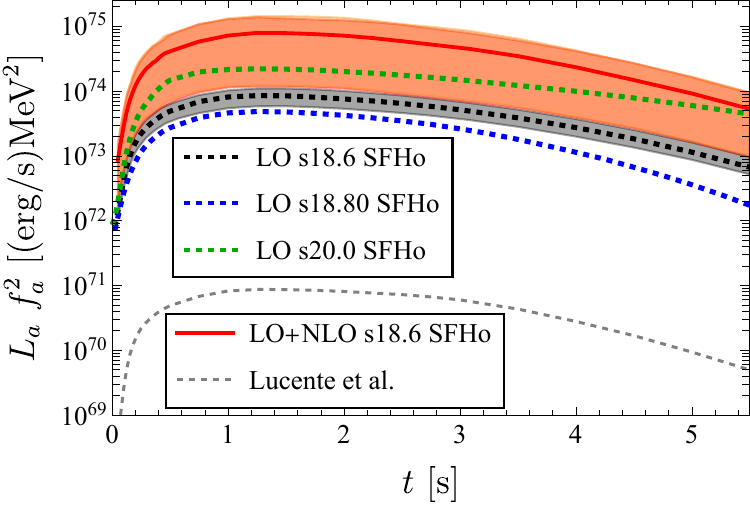}
\caption[]{Axion luminosity $L_a f_a^2$ as a function of time for different contributions as shown in \Fig{fig:Lumi}. The dashed lines show the LO emissivities, resulting from the diagram shown in \Fig{fig:c5_SN_diag}, for different supernova models.}
\label{fig:Lumi_model_dep}
\end{figure}

The supernova profiles resulting from a $18.8 \,M_\odot$ and a $20\,M_\odot$ progenitor from different simulations lead to a lower and a higher temperature profile, respectively.
The resulting temperature profiles for the three simulations as a function of the radius at $t_{pb}=1 \,\mathrm{s}$ are shown in \Fig{fig:profile_model_dep}.
Calculating the resulting axion luminosity as presented in the main text at LO, we find the results shown in \Fig{fig:Lumi_model_dep}.

While the results for the individual supernovae are quite different, especially at late times, we still find for all of them a strengthening of the supernova bound by around two orders of magnitude compared to~\cite{Lucente:2022vuo}.
The axion luminosity from the interaction taken into account in~\cite{Lucente:2022vuo} is only shown for the $18.6\,M_\odot$ progenitor mass. 
Since the relative effect from changing the SN model is identical to the change in luminosity with our interaction (blue, green dashed lines compared to black dashed line), they are suppressed in \Fig{fig:Lumi_model_dep}.
The supernova with slightly higher temperatures, where the progenitor has a mass of $20\,M_\odot$, leads to a strengthening of the LO bound by a factor of around $2.7$, while the supernova with slightly lower temperatures and a progenitor of a mass of $18.8\,M_\odot$, leads to a weakening of the LO bound by a factor of around $0.6$. Including also the NLO contributions and performing the same statistical analysis as above, similar results are to be expected.

\bibliography{bibliography}
\bibliographystyle{apsrev4-1}

\end{document}